\documentclass[conference]{IEEEtran}
\usepackage[numbers]{natbib} 
\IEEEoverridecommandlockouts
\usepackage{stfloats}
\usepackage{graphicx}  
\usepackage{amssymb}
\usepackage{epstopdf}
\usepackage{bm}
\usepackage{amsmath}
\usepackage{booktabs} 
\usepackage{multirow} 
\usepackage{makecell} 
\usepackage[table]{xcolor} 
\usepackage{float}   
\usepackage{subfigure} 
\usepackage{hyperref}
\usepackage{lipsum}
\hypersetup{hypertex=true,
colorlinks=true,
linkcolor=blue,
anchorcolor=blue,
citecolor=blue}
\usepackage[utf8]{inputenc}
\usepackage{booktabs}   
\usepackage{multirow}   
\usepackage{makecell}   
\usepackage[table]{xcolor} 
\usepackage{geometry}
\usepackage{tabularx}   
\usepackage[numbers]{natbib} 

\begin{document}

\title{RadioRange: An Open-Source Digital Twin-based Ranging Simulator for UWB, Wi-Fi, and 5G\\

}


\author{
\IEEEauthorblockN{1\textsuperscript{st} Zhen Lyu}
\IEEEauthorblockA{\textit{Department of Aeronautical and Aviation Engineering} \\
\textit{The Hong Kong Polytechnic University}\\Hong Kong, China\\
zhenn.lyu@connect.polyu.hk}
\and
\IEEEauthorblockN{2\textsuperscript{nd} Yidi Chen}
\IEEEauthorblockA{\textit{Department of Aeronautical and Aviation Engineering} \\
\textit{The Hong Kong Polytechnic University}\\Hong Kong, China\\
yi-di.chen@connect.polyu.hk}
\and
\IEEEauthorblockN{3\textsuperscript{rd} Li-ta Hsu}
\IEEEauthorblockA{\textit{Department of Aeronautical and Aviation Engineering} \\
\textit{The Hong Kong Polytechnic University}\\Hong Kong, China\\
lt.hus@polyu.edu.hk}
\and
\IEEEauthorblockN{4\textsuperscript{th} Guohao Zhang}
\IEEEauthorblockA{\textit{Department of Aeronautical and Aviation Engineering} \\
\textit{The Hong Kong Polytechnic University}\\Hong Kong, China\\
gh.zhang@polyu.edu.hk}
}

\maketitle

\begin{abstract}
Accurate RF-based ranging is critical for location-aware wireless systems, yet no open platform exists for fair, reproducible comparison across protocols under realistic hardware impairments. Existing simulators target communication-layer metrics and lack ranging algorithms, impairment models, and positioning-specific evaluation. We present RadioRange, an open-source, positioning-first digital twin that unifies UWB, Wi-Fi, and 5G~NR on identical ray-traced physical channels. The platform models eleven independently toggleable hardware impairments across three injection stages, spanning antenna-level offsets, RF circuit non-idealities, and post-compensation CSI residuals, each with documented physical models and protocol-specific defaults. Five first-path ranging detectors and three multipath identification algorithms are provided within a protocol-specific evaluation framework, enabling controlled Monte Carlo benchmarking and systematic ablation studies. The simulator is validated against real-world UWB and Wi-Fi measurements, demonstrating that the channel model faithfully captures geometry-dependent multipath bias. RadioRange-Sim is publicly available at \texttt{https://github.com/Togure/RadioRange}.
\end{abstract}

\begin{IEEEkeywords}
Indoor positioning, Observability, Radio SLAM, Extended Kalman Filter, Channel Impulse Response.
\end{IEEEkeywords}

\section{Introduction}
Accurate Radio-Frequency (RF) ranging is foundational to location-based services (LBS) and localization systems, including autonomous navigation, asset tracking, and the industrial IoT. Three wireless protocols currently dominate this landscape: UWB (IEEE 802.15.4z) offers centimeter-level accuracy through nanosecond pulse timing \cite{uwb_survey}; Wi-Fi (IEEE 802.11mc) provides fine-time measurement (FTM) on ubiquitous infrastructure \cite{lyu2025wi, wifi_rtt}; and 5G NR (FR1) enables wide-area positioning via dense pilot structures \cite{5g_positioning, 3gpp38901}. Each protocol employs different carrier frequencies, bandwidths, and signal structures, resulting in varying multipath resolution and hardware impairment susceptibility. Developing robust positioning systems, such as robotic sensor fusion, indoor fingerprinting, and Internet of Vehicles (IoV) applications, requires accurate modeling of signal propagation paths and ranging estimation. Currently, the community still lacks a unified platform to benchmark algorithms, compare RF protocols, and conduct accurate ranging simulations in real-world environments. This is mainly due to the following reasons:

First, most published studies on ranging rely on specific protocols. UWB papers evaluate leading-edge detection (LDE) algorithms against empirical UWB measurement datasets \cite{uwb_survey, uwb_lde}. Meanwhile, Wi-Fi FTM/RTT studies often use statistical TGn/TGax channel models \cite{80211mc, wifi_rtt}, and 5G NR positioning evaluations rely on 3GPP TR 38.901 stochastic cluster models \cite{3gpp38901, 5g_positioning} that fail to capture geometrically deterministic reflections. Few existing frameworks place all three protocols on identical, ray-traced channels. Here, the same room geometry, TX–RX positions, and material properties produce protocol-specific channel impulse responses (CIRs). Therefore, it is hard to simultaneously assess the performance differences of UWB, Wi-Fi, and 5G in positioning systems.

Second, most existing RF simulators are telecom-oriented rather than ranging-oriented. Ranging is a precise timing and phase-estimation problem, rather than a data-decoding problem. Yet, standard tools like MATLAB's 5G Toolbox \cite{matlab5g} or ns-3 are designed to optimize communication-layer metrics such as Bit Error Rate (BER), Error Vector Magnitude (EVM), and throughput. Even state-of-the-art physics simulators like NVIDIA Sionna \cite{sionna,wang2026r} generate high-fidelity MIMO channel responses but stop at the physical propagation layer. They abstract away the baseband processing of ranging, such as frequency-domain CIR reconstruction, sub-sample interpolation, and LDE, which are the core of localization systems. Consequently, there is a distinct lack of simulators designed at the baseband signal level to evaluate ranging algorithms.

Finally, real-world ranging error is dominated by hardware non-idealities, which are absent from pure mathematical models. Even when ranging algorithms are evaluated, they often assume ideal transceivers and treat thermal noise as the sole source of error. 
In reality, ranging accuracy is severely degraded by hardware constraints: ADC quantization dithers CIR peaks \cite{dw1000}, I/Q gain and phase mismatch create mirror-image ghost paths that deceive first-path detectors \cite{iq_imbalance_wifi}, and sampling clock jitter randomly shifts CIR bins \cite{impairment_survey}. Because most current ranging platforms fail to inject these circuit- and CSI-level impairments before the algorithm layer, their simulated distance errors rarely match real-world trajectory measurements, leading to overly optimistic evaluations.

To address these challenges, we present {RadioRange} simulator, an open-source digital twin designed to unify RF ranging simulation and evaluation. Our platform's primary contributions are as follows:

\begin{itemize}
    \item evaluates UWB (499.2 MHz), Wi-Fi (160 MHz, 802.11az), and 5G NR (122.88 MHz, FR1 n79) on identical ray-traced channels \cite{sionna} through a CIR interface, enabling the cross-protocol comparison.

    \item design a multi-stage injection pipeline featuring 11 independently toggleable hardware impairments.  It ensures simulated ranging errors accurately capture real-world hardware constraints. 

    \item provides a standardized benchmark comprising five first-path detectors and three multipath identification algorithms. The simulator's accuracy is validated against UWB and Wi-Fi measurements along two real-world indoor trajectories.

\end{itemize}

\section{Simulator System Overview and Fundamental Principles}
\label{sec:overview}

\begin{figure*}[htbp]
  \centering
\includegraphics[width=0.95\linewidth]{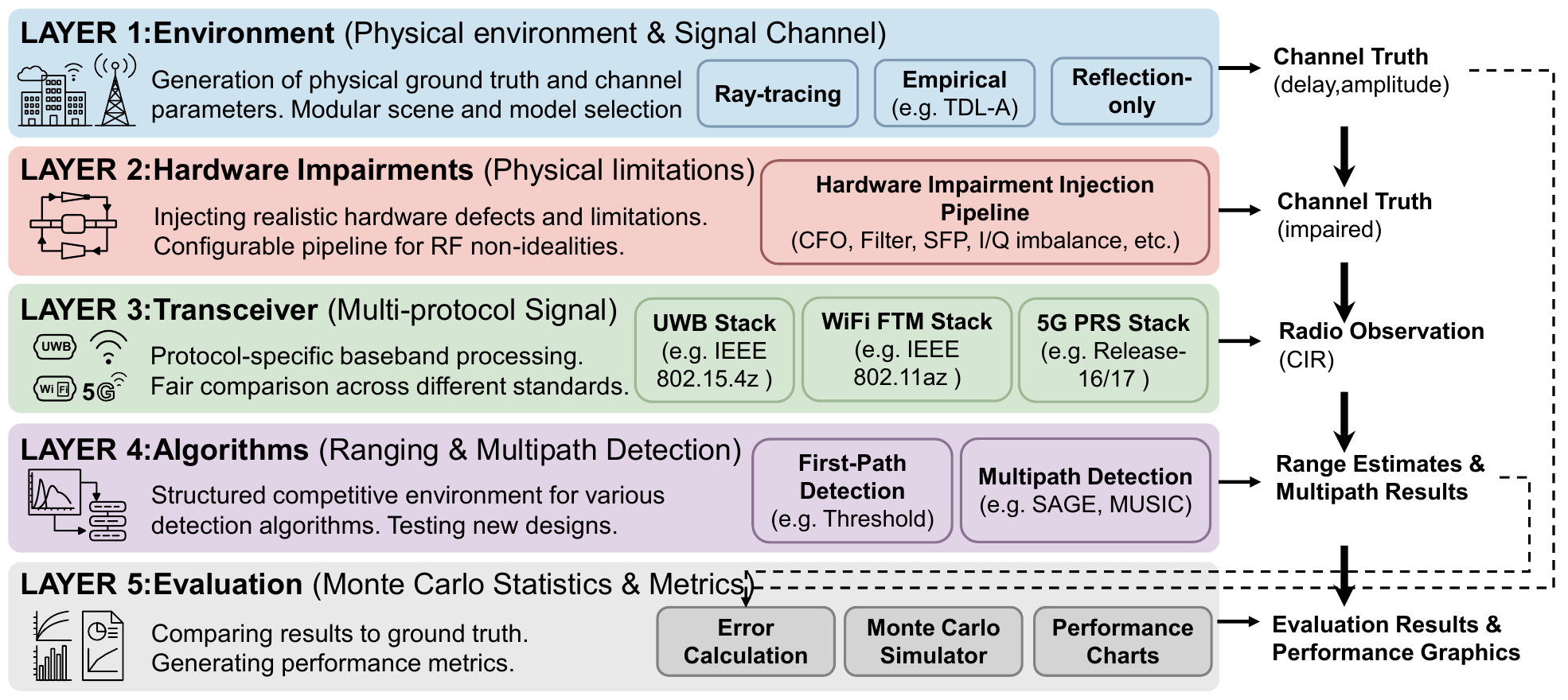}
  \caption{Five-layer architecture of RadioRange simulator.}
  \label{fig:architecture}
\end{figure*}

Fig.\ref{fig:architecture} illustrates the five-layer architecture of our simulator. Layer 1 (Environment) models the channel impulse response by generating propagation paths via Sionna \cite{sionna} ray-tracing or a reflection-only engine to yield a unified \texttt{ChannelTruth}. Layer 2 (Hardware Impairments) injects 11 independently toggleable physical and receiver-level errors (e.g., SFO, ADC quantization) to bridge the gap between theoretical channels and real hardware. Layer 3 (Transceiver) provides a common interface for UWB, Wi-Fi, and 5G to construct the frequency response, apply impairments, and compute the CIR, outputting a protocol-independent \texttt{RadioObservation}. Layer 4 (Algorithms) processes this observation without accessing protocol specifics or ground truth, producing a \texttt{RangeEstimate} or \texttt{MultipathResult}. Finally, Layer 5 (Evaluation) aggregates reproducible Monte Carlo statistics (e.g., RMSE, error CDF) across spatial trajectories or randomized trials.

\subsection{Ranging Fundamentals: UWB, Wi-Fi, and 5G}
\label{sec:ranging_fundamentals}

This subsection explains the difference between UWB, Wi-Fi, and 5G in ranging results.

\begin{table}[htbp]
  \centering
    \setlength{\tabcolsep}{1pt} 
  \caption{Protocol specifications for UWB, Wi-Fi, and 5G NR as default configured in RadioRange. res. is the abbreviation for resolution.}
  \label{tab:protocols}
  \small
  \begin{tabular}{@{}lccc@{}}
    \toprule
    \textbf{Parameter} & \textbf{UWB} & \textbf{Wi-Fi}
    & \textbf{5G NR} \\
    \midrule
    Protocol & IR-UWB&802.11mc&FR1\\
    Carrier frequency          & 7.99\,GHz (Ch~9)  & 5.18\,GHz        & 4.80\,GHz (n79) \\
    Bandwidth  & 499.2\,MHz        & 160\,MHz         & 122.88\,MHz \\
    FFT size / CIR bins        & 1024              & 512              & 4096 \\
    Subcarrier spacing         & --                & 312.5\,kHz       & 30\,kHz \\
    Spectral window            & Hamming           & None             & None \\
    ADC resolution             & 6\,bit  & 10\,bit          & 12\,bit \\
    Range res.\ per bin        & 0.60\,m           & 1.88\,m          & 2.44\,m \\
    Rayleigh $\Delta\tau$      & 2.0\,ns  & 6.3\,ns & 8.1\,ns\\
    \bottomrule
  \end{tabular}
  \vspace{-3pt}
\end{table}

The Channel Impulse Response (CIR) is the observable for all
three ranging protocols:
\begin{equation}
  h(t) = \sum\nolimits_{i} a_i \, \delta(t - \tau_i),
  \label{eq:cir}
\end{equation}
where $a_i$ is the complex path gain and $\tau_i$ the one-way
time of flight.  $\tau_0$ is the time delay of the first arriving path. 
Although their PHY layers operate differently, they are unified at the CIR level, enabling evaluation through an algorithm interface. UWB (499.2\,MHz bandwidth) directly extracts the CIR via time-domain matched filtering against a physical nanosecond pulse, achieving a fine Rayleigh resolution of $\Delta \tau \approx$2.0 ns (0.60 m per bin) that resolves dense multipath. 
In contrast, Wi-Fi and 5G perform OFDM channel estimation to find the frequency response $H(f_k)$, followed by an IFFT: $h(t)=IFFT ({H}(f_k))$.
This frequency-domain division and IFFT is equivalent to time-domain cross-correlation (as used in 802.11mc RTT first-peak detection). However, Wi-Fi and 5G construct $H(f_k)$ differently due to their distinct signal structures. 
Wi-Fi (160 MHz, 1.88 m resolution) derives its estimate from a Long Training Symbol (LTS) preamble that is densely and continuously populated across all active subcarriers, providing a single-shot, contiguous frequency snapshot. Conversely, 5G (122.88\,MHz, 2.44 m resolution) relies on Positioning Reference Signals (PRS) that are sparsely distributed in a comb-like grid across specific subcarriers. The 5G receiver must perform frequency-domain interpolation between these discrete pilots to reconstruct the full channel. While this sparse grid yields the coarsest temporal resolution, 5G trades pure bandwidth for robustness: its structured pilot grid enables superior Carrier Frequency Offset (CFO) tracking and channel estimation in low-SNR regimes compared to Wi-Fi.

\begin{figure}[htbp]
\centering\includegraphics[width=0.9\linewidth]{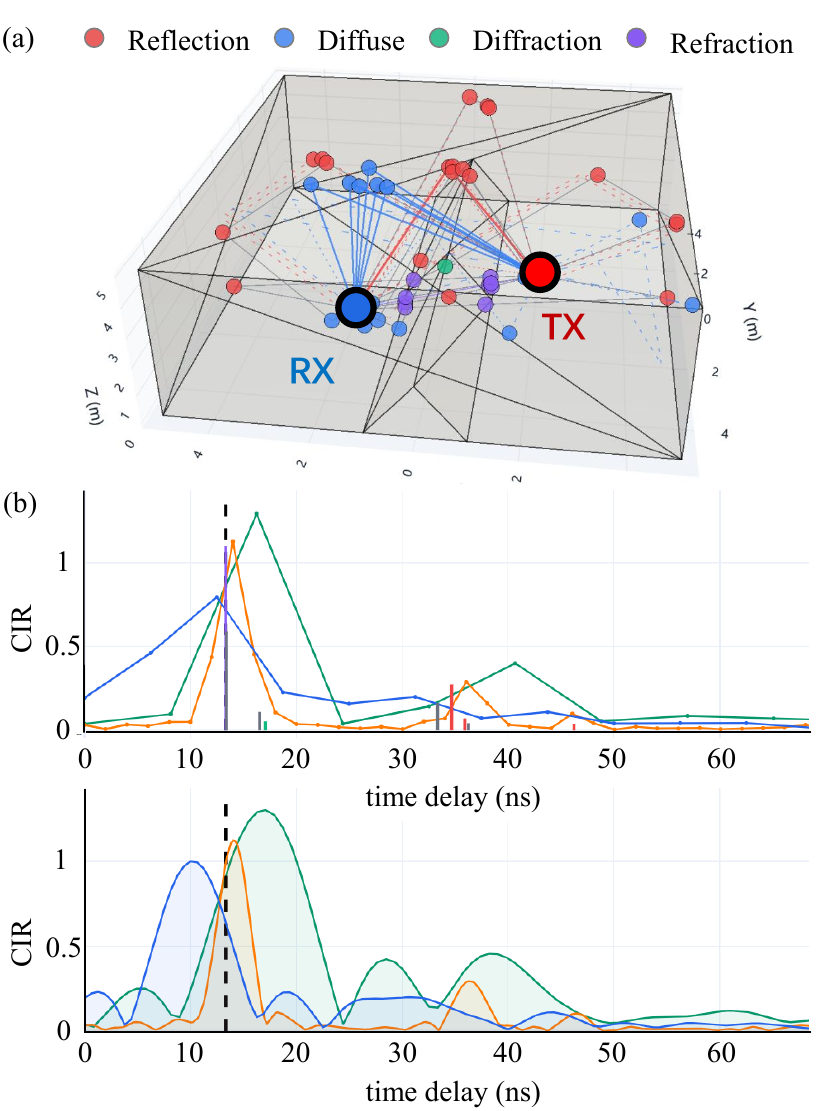}
    \caption{(a) Ray-tracing simulation in a box with the signal's NLOS propagation. (b) Corresponding discrete CIR and continuous CIR.}
    \label{fig:Figure3}
\end{figure}

\subsection{Channel Propagation Modeling}
\label{sec:channel_modeling}

To generate the ground-truth propagation paths (complex gain $a_i$, delay $\tau_i$), RadioRange employs a dual-engine architecture. Both engines output identical \texttt{ChannelTruth} structures, ensuring the downstream pipeline remains engine-agnostic.

\begin{figure}[htbp]
    \centering
\includegraphics[width=1\linewidth]{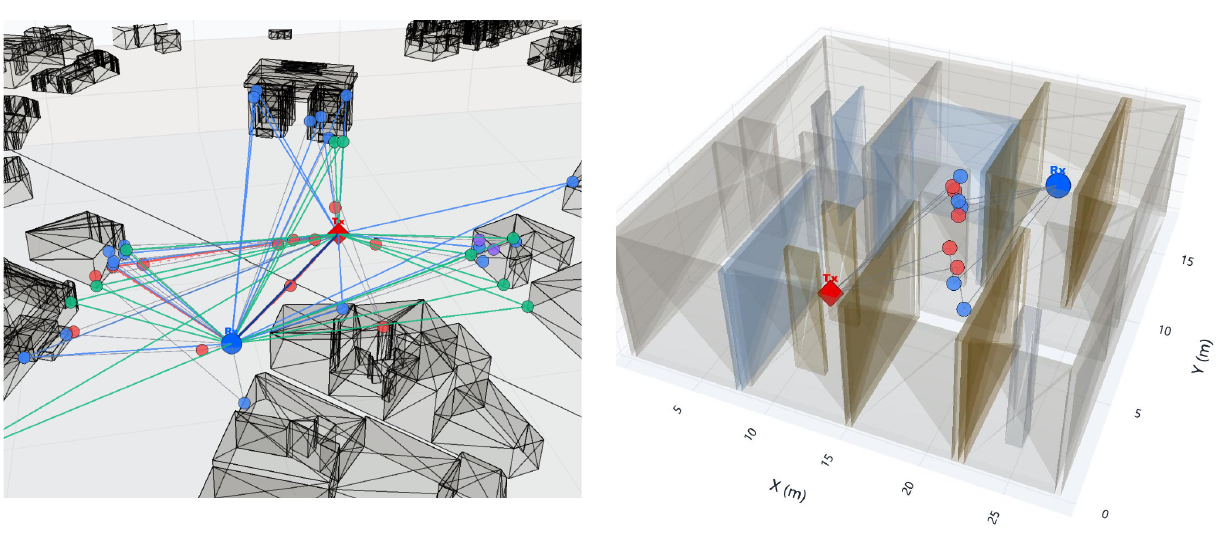}
    \caption{Sionna-based ray-tracing simulation in the Triumphal arch and self-defined environment.}
    \label{fig:figure2}
\end{figure}

\textbf{High-Fidelity Ray Tracing (Sionna RT).} We leverage NVIDIA Sionna's Shooting-and-Bouncing Rays (SBR) engine for physically accurate 3D scene modeling. Rays are traced with up to \texttt{max\_depth} bounces, capturing line-of-sight, specular reflection, diffuse scattering, refraction, and diffraction. Scene materials are defined by relative permittivity $\varepsilon'$ and conductivity $\sigma$ per ITU-R P.2040 recommendations. Crucially, Sionna computes frequency-dependent Fresnel reflection coefficients at each protocol's specific carrier frequency (UWB @8.0\,GHz, Wi-Fi @5.2\,GHz, 5G @4.8\,GHz). Re-running the ray tracer per band (rather than simply frequency-scaling a single run) ensures physically correct path gains, as materials like concrete ($\varepsilon' = 5.31$) and metal ($\sigma = 10^7$\,S/m) exhibit dispersive reflection properties. To maintain efficiency, these computationally intensive RT results are cached and reused across all downstream impairment and algorithm sweeps. Fig.\ref{fig:figure2} shows the Sionna-based ray-tracing simulation illustration.

\textbf{Reflection-only image method.} For high-throughput statistical evaluation, we implement a geometric image-method engine. It analytically computes specular reflections for rectangular rooms up to a specified bounce order in $<1$\,ms per trial. While it omits diffuse scattering and diffraction, its extreme speed and geometric determinism make it ideal for massive parameter sweeps (e.g., hundreds of trials in seconds) and trajectory-based validation against physical rectangular environments.

\textbf{Statistical Channels.} Beyond geometric modeling, the platform supports 3GPP TDL-A through TDL-E statistical channel models, deterministic two-path scenarios, and manual path definitions for debugging and pedagogical purposes.

\IEEEpeerreviewmaketitle


\subsection{Hardware Impairment Modeling and Ablation}
\label{sec:impairments}
Hardware impairments are injected at three stages, distinguished by
whether the error is applied before or after the
transceiver's DSP compensation chain:

\textbf{Channel-level (pre-DSP):} Impairments that modify the raw
physical path parameters in \texttt{ChannelTruth} before the
transceiver observes them.  These represent physical errors present
at the antenna interface, before any digital compensation.

\textbf{Circuit-level (HW):} Impairments introduced by the
analog/RF hardware during CIR construction, including ADC quantization,
automatic gain control, and I/Q down-converter mismatch.

\textbf{CSI-level (post-DSP):} Frequency-domain residual errors
applied during OFDM channel estimation, after pilot-based
compensation.  These are the errors that remain after
the receiver's carrier recovery, SFO tracking, and phase correction
loops have operated.




\textbf{ADC Quantization (UWB Vulnerability):} 
Coarse ADC resolution severely distorts UWB signals, causing a 22.6\% drop in ranging accuracy. Because UWB relies on precise time-domain amplitude tracking for its leading-edge detection, aggressive quantization degrades the fidelity of the pulse envelope. In contrast, Wi-Fi and 5G remain largely unaffected ($<\!1\%$), as their OFDM-based ranging accuracy is primarily bandwidth-limited rather than amplitude-limited.

\textbf{I/Q Imbalance (Wi-Fi/5G Vulnerability):} 
Gain and phase mismatches between the I and Q branches cause the most significant degradation for Wi-Fi ($-18.0\%$) and 5G ($-4.6\%$). This mismatch creates a mirror image $\hat{H}^*(-f_k)$ in the frequency domain, introducing a spurious peak in the CIR envelope that threshold-based algorithms often misidentify as the first path. UWB is highly resilient to this ($<\!2.5\%$), as its leading-edge detector operates on the initial rising edge before mirror-induced envelope distortions appear.

\textbf{Minor Effects and DSP Mitigation:} 
ADC timing jitter introduces only minor degradation for Wi-Fi ($-2.1\%$) and 5G ($-3.8\%$). Crucially, 8 of the 11 modeled impairments (e.g., CFO, SFO, Channel Estimation) show negligible impact ($<\!1\%$). This provides quantitative evidence that modern receiver DSP—such as per-subcarrier pilot tracking and PLL carrier recovery—successfully suppresses these raw physical errors below the indoor multipath noise floor. Ultimately, when all 11 impairments are active, the aggregate accuracy drops ($-28.6\%$ for UWB, $-23.0\%$ for Wi-Fi, and $-9.1\%$ for 5G) are driven almost entirely by the uncompensated hardware-level distortions (ADC limits and I/Q mismatch).


\section{Ranging and Multipath Algorithms}
\label{sec:algorithms}

Given a CIR, this section describes the shared signal processing infrastructure and the specific algorithms evaluated for first-path ranging and multipath identification. For multipath detection, we integrate standard algorithms including PeakFinder, CA-CFAR \cite{gandhi1988analysis}, and CLEAN \cite{cramer2002evaluation}. Detailed mathematical derivations are omitted due to space constraints, but their configurations and comparative principles are summarized in Table~\ref{tab:combined_algos}.

\begin{table}[htbp]
  \centering
  \caption{Summary of evaluated ranging and multipath identification algorithms.}
  \label{tab:combined_algos}
  \small
  \setlength{\tabcolsep}{6pt} 
  \begin{tabular}{@{}lp{6.5cm}@{}}
    \toprule
    \textbf{Algorithm} & \textbf{Principle} \\
    \midrule
    \multicolumn{2}{@{}l}{\textbf{First-Path Ranging Algorithms}} \\
    \midrule
    MaxPeak
      & $\hat{\tau}_0 = \arg\max |\text{CIR}(t)|$. The naive baseline finding the global maximum. \\
    Threshold
      & Scans forward to find the first bin exceeding a fixed relative threshold (e.g., $0.18 \cdot \max(|\text{CIR}|)$). \\
    LeadingEdge
      & Uses an adaptive noise floor ($\mu + 4\sigma$) and requires a minimum run of consecutive bins above the threshold to reject isolated spikes. \\
    SearchBack
      & Scans backward from the global peak to find the first bin that drops below a relative threshold, locating the rising edge. \\
    ChipLDE
      & Mimics DW1000/DW3000 hardware logic using a fixed offset ($+10$\,dB) above the noise floor coupled with a consecutive-bin persistence check. \\
    \midrule
    \multicolumn{2}{@{}l}{\textbf{Multipath Identification Algorithms}} \\
    \midrule
    PeakFinder
      & Identifies local maxima in the CIR envelope that exceed a configured dynamic range above the noise floor. \\
    CA-CFAR
      & Cell-Averaging Constant False-Alarm Rate; adapts a per-bin threshold based on the mean power of surrounding training cells. \\
    CLEAN
      & Complex-domain iterative deconvolution; recursively finds the strongest peak and subtracts a complex sinc template to resolve sub-Rayleigh paths. \\
    \bottomrule
  \end{tabular}
\end{table}

\subsection{From CIR to Distance}
\label{sec:pipeline}

All algorithms consume a protocol-independent \texttt{RadioObservation} and share two critical preprocessing steps:
\begin{enumerate}
    \item \textbf{Sub-sample refinement:} Algorithms first identify a coarse peak, then map to a 10$\times$ zero-padded continuous CIR (equivalent to ideal band-limited sinc interpolation) to perform a local peak search, yielding decimeter-level resolution.
    \item \textbf{Noise floor estimation:} The noise statistics ($\mu_{\text{noise}}$, $\sigma_{\text{noise}}$) are dynamically estimated from the CIR tail (last 30\% of bins). This shared baseline ($\mu + 3\sigma$) ensures robust thresholding across varying SNRs without manual, per-scene calibration.
\end{enumerate}

\subsection{First-Path Ranging Algorithms}
\label{sec:first_path}

Rather than duplicating their exact mathematical mechanisms, we highlight the comparative robustness of the five first-path detectors:

\textbf{MaxPeak} serves as a naive baseline that fails catastrophically in NLOS conditions or when specular reflections are stronger than the direct path. \textbf{Threshold} and \textbf{SearchBack} rely on a fixed relative threshold ($\alpha=0.18$) and perform well in simple environments.

\section{Experimental Evaluation}
\label{sec:evaluation}
We conducted a three-layer experiment. First, we compared the performance of UWB ( Ch~9, 7.99\,GHz, 499.2\,MHz, 1024 bins), Wi-Fi (802.11 OFDM, 5.18\,GHz, 160\,MHz, 512 subcarriers), and 5G NR (OFDM, 4.80\,GHz, 122.88\,MHz, 4096 subcarriers) under the same physical channel but with different reflective surface material configurations (concrete, glass, and metal) to determine the raw signal quality upon which downstream algorithms rely. Second, we benchmarked all five first-path ranging algorithms under three protocols and two impairment conditions, reporting ranging accuracy. Third, we validated the simulator's accuracy using real UWB and Wi-Fi ranging measurement data collected from two indoor tracks.

\subsection{Simulation Setup}
\label{sec:setup}

 All experiments use NVIDIA Sionna's Shooting-and-Bouncing Rays (SBR) engine for physically accurate path generation.  For the Monte Carlo test, the box room with an obstacle scene is used with 300 random TX-RX placements per protocol and algorithm configuration. Three material variants are evaluated: all-concrete materials,  concrete walls and glass obstacle, and all-metal materials. For real-data validation, we replicate two real-world measurement campaigns: a 25\,m straight corridor and a L-junction.  The simulator uses identical room geometry, wall materials, and TX-RX coordinates as the real measurements.

\subsection{CIR-Level Protocol Comparison}
\label{sec:cir_comparison}

\begin{figure}[htbp]
  \centering
   \includegraphics[width=0.9\linewidth]{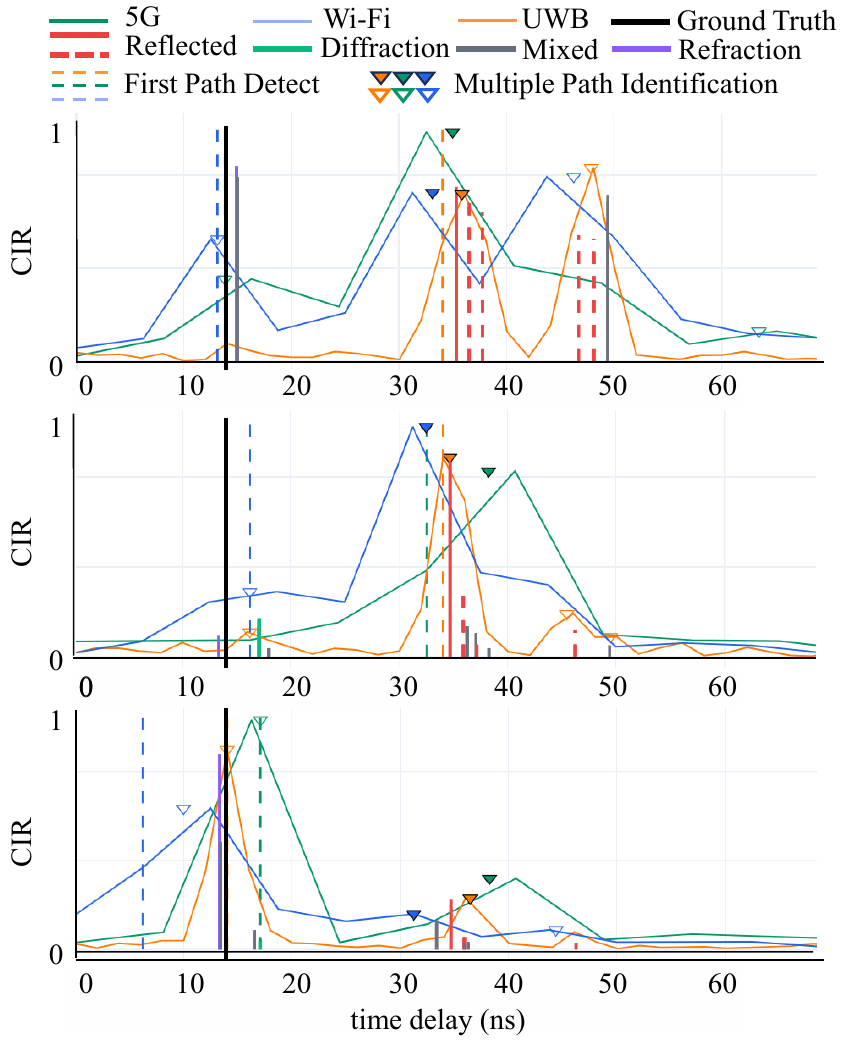}
  \caption{CIR curves for UWB, Wi-Fi, and 5G observed on different material configurations of the NLOS physical channel (corresponding to all-concrete, all-metal, and concrete wall with glass obstruction, respectively). The vertical lines are the true values of different propagation paths.}
  \label{fig:cir_comparison}
\end{figure}
\begin{figure}[htbp]
  \centering
   \includegraphics[width=0.9\linewidth]{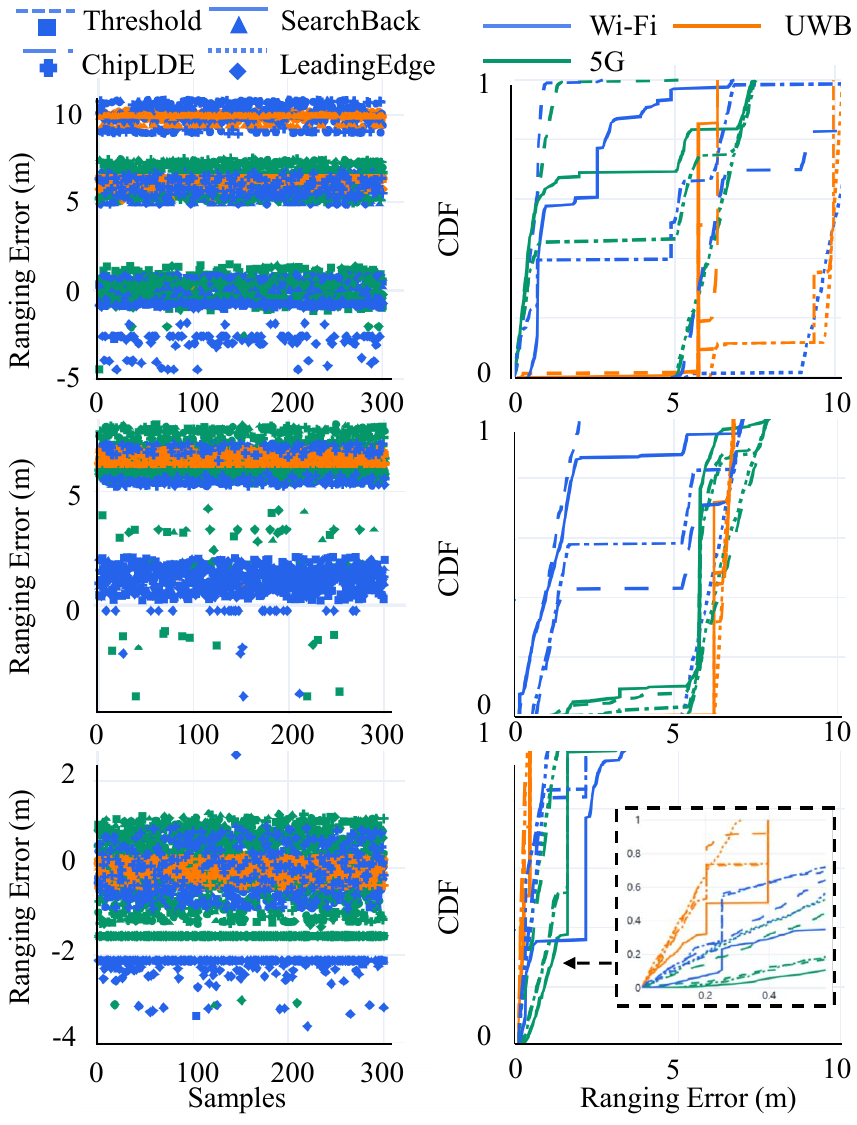}
  \caption{Ranging error CDF multi-panel comparison.
           Top-left: protocol-level comparison (best algorithm per
           protocol, no impairments).
           Top-right: protocol-level comparison under full impairments.
           Bottom-left: algorithm-level comparison for UWB
           (Threshold, SearchBack, ChipLDE, LeadingEdge).
           Bottom-right: algorithm-level comparison for Wi-Fi.
           Inset zoom boxes magnify the P10–P90 region.
           The protocol ranking (UWB $>$ Wi-Fi $>$ 5G) and algorithm
           ranking (ChipLDE $\approx$ LeadingEdge $>$ Threshold $>$
           SearchBack $>$ MaxPeak) are visible in a single glance.}
  \label{fig:cdf_comparison}
\end{figure}

 Before evaluating ranging accuracy, we first examine the raw channel observable  CIR.  Fig.~\ref{fig:cir_comparison} overlays the UWB, Wi-Fi, and 5G observed CIRs on a common time-delay axis for one NLOS trial in the concrete box room.   UWB cleanly separates multipath components that Wi-Fi and 5G merge into single broadened peaks. We compare three wall-material configurations (metal, all-concrete, and concrete wall and glass obstacles).  The CIR values of the three different materials differ significantly. In an all-metal environment, the attenuation of the direct path is severe, while the attenuation of the reflected path is weaker, with the reflected path becoming the dominant peak. In an all-concrete environment, the attenuation of the direct path is still severe, and the primary reflection is relatively significant, but the secondary reflection intensity is weaker compared to the all-metal environment. This phenomenon is more pronounced in UWB (UWB operates at a higher frequency). In glass obstruction and concrete reflection, the attenuation of the direct path is smaller, and it is stronger relative to the reflection, therefore the primary diameter is near the direct path, resulting in a smaller ranging error. Fig.~\ref{fig:cdf_comparison} presents the error CDFs as a 4-panel
composite, providing both protocol-level and algorithm-level
comparison.  The UWB CDF rises sharply: 90\% of errors are below
0.38\,m (P90) under no impairments for the best algorithm in the
concrete box scene.  Wi-Fi and 5G CDFs rise more gradually, with P90
values of 0.84\,m and 1.09\,m respectively.  The inset zoom boxes
in the P10–P90 region enable fine discrimination between closely-
performing algorithms (e.g., ChipLDE vs.\ LeadingEdge for UWB).

\subsection{Ranging Algorithm Benchmark}
\label{sec:algo_benchmark}
\begin{table}[htbp]
  \centering
  \caption{Ranging RMSE (m) for all 5 first-path algorithms across
           UWB, Wi-Fi, and 5G in different scenes.}
  \label{tab:rmse_matrix}
  \small
  \setlength{\tabcolsep}{4.5pt} 
  \begin{tabular}{@{}l ccc | ccc@{}}
    \toprule
    & \multicolumn{3}{c}{\textbf{Box Room}}
    & \multicolumn{3}{c}{\textbf{Corridor}} \\
    \cmidrule(lr){2-4} \cmidrule(l){5-7}
    \textbf{Algorithm}
      & \textbf{UWB} & \textbf{Wi-Fi} & \textbf{5G}
      & \textbf{UWB} & \textbf{Wi-Fi} & \textbf{5G} \\
    \midrule
    MaxPeak
      & 0.19 & 0.57 & \textbf{0.67}
      & 0.99 & 0.99 & 0.97 \\
    Threshold ($\alpha{=}0.18$)
      & 0.24 & 0.94 & 1.22
      & 0.28 & 1.44 & 2.80 \\
    LeadingEdge ($4\sigma$)
      & 0.30 & 1.85 & 1.36
      & \textbf{0.25} & 3.76 & 3.39 \\
    SearchBack ($\alpha{=}0.18$)
      & 0.24 & 0.86 & 1.24
      & 0.97 & 1.44 & 2.69 \\
    ChipLDE (10\,dB)
      & \textbf{0.18} & \textbf{0.53} & 0.72
      & 0.39 & \textbf{0.89} & \textbf{0.85} \\
    \bottomrule
  \end{tabular}
  \vspace{4pt}
  \footnotesize
  \vspace{-6pt}
\end{table}

\begin{figure}[htbp]
  \centering
   \includegraphics[width=\columnwidth]{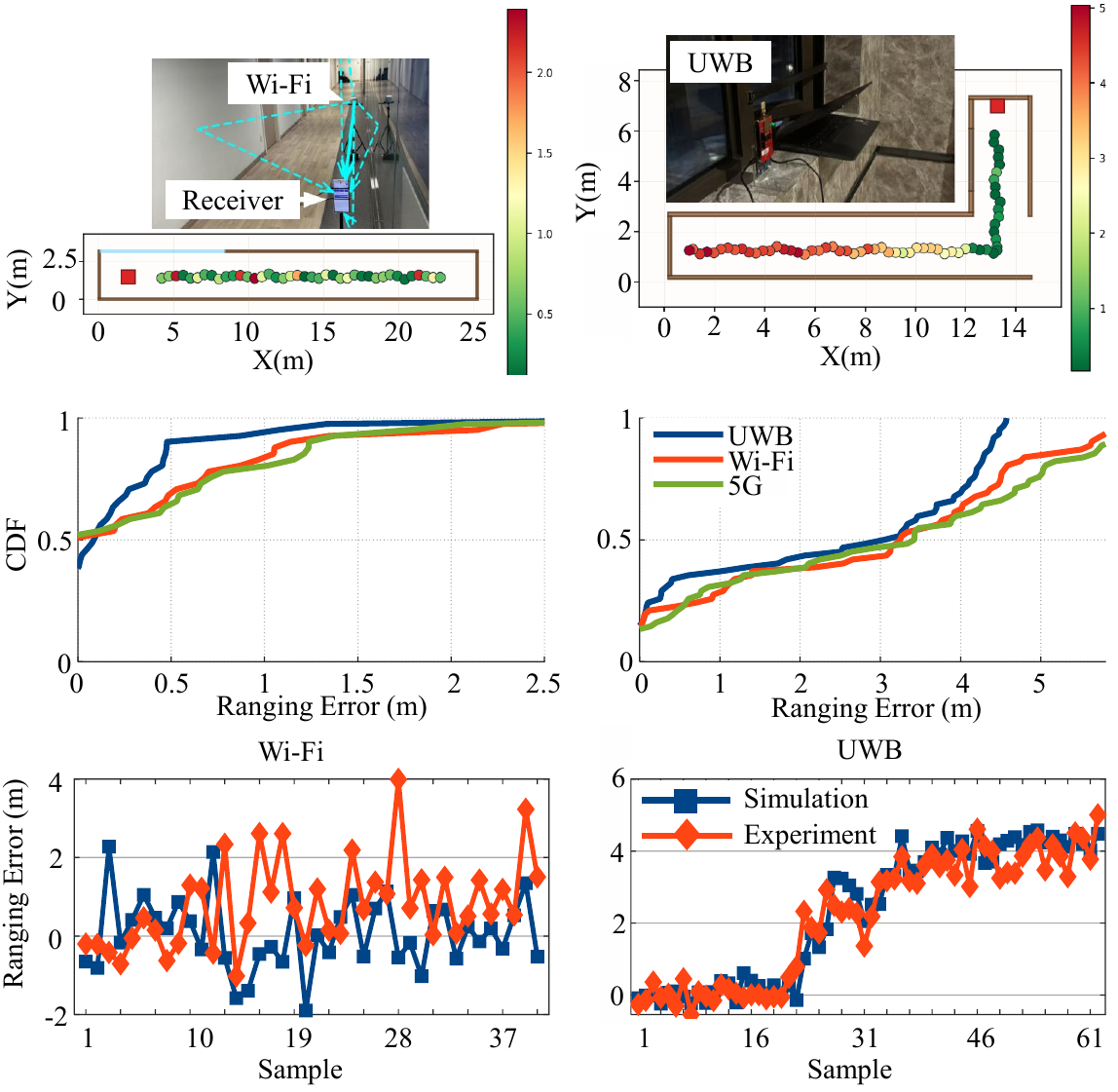}
  \caption{Top row: Experimental setups and trajectory ranging error heatmaps for Wi-Fi (corridor) and UWB (L-junction). Middle row: CDF of ranging errors comparing UWB, Wi-Fi, and 5G protocols. Bottom row: Per-sample error traces overlaid for simulation and experiment.}
  \label{fig:validation}
\end{figure}

Table~\ref{tab:rmse_matrix} reports the RMSE of all five first-path algorithms across the three protocols for two representative scenes: a simple box room and a 25\,m straight corridor (shown in Fig.\ref{fig:validation}).  Attributable to its finer resolution, UWB consistently outperforms Wi-Fi and 5G by a factor of 2–4× in RMSE, with its best algorithm (ChipLDE, 0.19 m) proving 2.8× and 3.8× more accurate than the best Wi-Fi and 5G counterparts under ideal conditions. However, this high resolution can paradoxically make UWB more vulnerable in dense multipath environments, where strong, distinct reflections may be misidentified as the first path, whereas Wi-Fi might merge them into a single peak. 
Across all protocols, ChipLDE emerges as the most robust algorithm, achieving the lowest RMSE in five of six scenarios by utilizing a hardware-mimicking fixed threshold (+10 dB) and a persistence check to reliably reject noise and detect the true first path. In contrast, LeadingEdge (4$\sigma$) performs well for UWB but degrades severely for Wi-Fi and 5G (up to 3.76 m RMSE), as their broader CIR peaks artificially inflate the adaptive noise-floor threshold and cause missed detections. Finally, MaxPeak proves inadequate for anything but pure line-of-sight (LOS) conditions, failing catastrophically whenever subsequent reflections exceed the direct path's amplitude due to physical obstructions or material attenuation.

\subsection{Real-World Trajectory Validation}
\label{sec:validation}

To validate the simulator's feasibility, we compared simulated ranging errors against real-world measurements using QM33120WDK1 UWB kit and Google Nest Wi-Fi hardware based on 802.11mc. The experiments were conducted along two indoor trajectories: a 25m straight corridor (line-of-sight dominant) and a challenging L-junction (non-line-of-sight dominant).The simulation accurately reproduces real-world behavior in both overall distribution and localized fluctuations. In the NLOS L-junction, the simulated RMSE effectively mirrors the real-world tripling of error caused by severe multipath bias. More importantly, the simulator demonstrates high per-sample tracking fidelity. By overlaying the simulated and measured error traces, we observe that the simulation tracks experimental fluctuations sample-by-sample, accurately capturing the geometry-dependent spatial variation of multipath effects. The simulator captured 85–90\% of the true ranging-error variance for both UWB and Wi-Fi.

\section{Conclusion}
\label{sec:conclusion}

We introduced RadioRange, an open-source digital twin that unifies UWB, Wi-Fi, and 5G NR ranging evaluation over identical ray-traced channels with 11 toggleable hardware impairments and a structured algorithm arena.
Our comprehensive evaluation reveals that:
1.UWB's superior bandwidth provides a 2--4$\times$ RMSE advantage over Wi-Fi and 5G under identical conditions (0.19\,m vs.\ 0.53\,m vs.\ 0.72\,m in simple-LOS).
2.Modern DSP effectively neutralizes 8 of the 11 modeled impairments to below a 1\% impact; only I/Q imbalance, ADC timing jitter, and ADC quantization (which uniquely \emph{improves} UWB via stochastic resonance) significantly alter ranging accuracy.
Real-world trajectory validation (102 samples) confirms the simulator's high fidelity with sub-4,cm mean bias for UWB. Future work will expand to dynamic channel geometries, ML-based detectors, and super-resolution algorithms (MUSIC/ESPRIT). RadioRange is open-source at \texttt{https://github.com/Togure/RadioRange}.

\small
\bibliography{IEEEabrv,reference}



\end{document}